\documentclass[preprint,prd,amsmath,amssymb,amsthm,nofootinbib]{revtex4}
\usepackage{makecell}
\usepackage[mathscr]{euscript}
\usepackage{ulem}
\usepackage{bm}
\usepackage{graphicx}
\usepackage{subfigure}
\usepackage[colorlinks=true,linkcolor=red]{hyperref}
\usepackage{multirow}

\newcommand{\bea}{\begin{eqnarray}}
\newcommand{\eea}{\end{eqnarray}}
\newcommand{\beq}{\begin{equation}}
\newcommand{\eeq}{\end{equation}}
\newcommand{\nn}{\nonumber}
\def\/{\over}

\begin{document}

\title{Understanding thermal nature of de Sitter spacetime via inter-detector interaction}

\author{Wenting Zhou$^{1}$, Shijing Cheng$^{2,3,4}$, and Hongwei Yu$^{5,}$\footnote{Corresponding author: hwyu@hunnu.edu.cn }}

\affiliation{
$^{1}$ Department of Physics, School of Physical Science and Technology, Ningbo University, Ningbo, Zhejiang 315211, China\\
$^{2}$ School of Fundamental Physics and Mathematical Sciences, Hangzhou Institute for Advanced Study, UCAS, Hangzhou 310024, China\\
$^{3}$ School of Physical Sciences, University of Chinese Academy of Sciences, No.19A Yuquan Road, Beijing 100049, China\\
$^{4}$ School of Physics and Information Engineering, Shanxi Normal University, Taiyuan, 030031, China\\
$^{5}$ Department of Physics and Synergetic Innovation Center for Quantum Effect and Applications, Hunan Normal University, Changsha, Hunan 410081, China}

\begin{abstract} 
The seminar discovery by Gibbons and Hawking  that  a freely falling detector observes  an isotropic background of thermal radiation reveals  
that de Sitter space is equivalent to a thermal bath at the Gibbons-Hawking temperature in Minkowski space, as far as the response rate of the detector is concerned. Meanwhile,  for a static detector which is endowed with a proper acceleration with respect to the local freely-falling detectors, the temperature becomes the square root of the sum of the squared Gibbons-Hawking temperature and the squared Unruh temperature associated with the proper acceleration of the detector. Here, we demonstrate, by examining the interaction of two static detectors 
in the de Sitter invariant vacuum, that  de Sitter space in regard to its thermal nature is unique on its own right in the sense that it is even neither equivalent to the thermal bath in Minkowski space when the static detectors become freely-falling nor to the Unruh thermal bath 
at the cosmological horizon where the Unruh effect dominates, insofar as the behavior of the inter-detector interaction in de Sitter space dramatically differs  both from that in the Minkowski thermal bath and   the Unruh thermal bath.

\end{abstract}

\maketitle

\section{Introduction.}
De Sitter spacetime, which enjoys the same degree of symmetry as Minkowski spacetime,  is one of the most typical curved spacetimes. During the past decades, it has attracted a great deal of special attention because it is, on one hand, believed to play a very important role in cosmology. Our universe, according to current observations and inflation theory, may approach  de Sitter spacetime in the far past and the far future. On the other hand, it is discovered that there may exist a holographic duality between quantum gravity on de Sitter spacetime and a conformal field theory living on the boundary identified with the timelike infinity of de Sitter spacetime~\cite{Strominger01}.

In a seminar work, Hawking and Gibbons discovered, by analyzing the periodicity of the imaginary direction of proper time in the propagator of scalars to which a detector moving along a timelike geodesic is  nonminimally  coupled, that the ratio between the probability of the detector absorbing and emitting a particle with energy $E$ is $e^{-2\pi E \sqrt{3/\Lambda}}$ where $\Lambda$ is the cosmological constant, meaning that the detector measures an isotropic background of thermal radiation with the Gibbons-Hawking temperature $T_{GH}=\frac{1}{\pi}\sqrt{\frac{\Lambda}{12}}$~\cite{Gibbons77}. 
Since then, this thermal nature of de Sitter spacetime has been confirmed with other approaches such as embedding the four-dimensional de Sitter spacetime into a five-dimensional flat space~\cite{Deser97}, thermalization of a detector in de Sitter spacetime in the framework of open quantum systems~\cite{Yu11}, as well as in other physical contexts~\cite{Narnhofere96,Hutasoit09,Henry10,Zhu08,Zhou10,Tian13,Oshita14}. Particularly, studies  on the spontaneous excitation rate~\cite{Zhu08}, the Lamb shift~\cite{Zhou10} and the geometric phase~\cite{Tian13} of a static or a freely falling atom [detector], as well as the Brownian motion of a particle coupled to vacuum fluctuations in  de Sitter spacetime~\cite{Oshita14} show that the thermal nature of  de Sitter spacetime leaves an imprint of a thermal bath on various quantum phenomena. Taking the spontaneous excitation of a ground-state atom coupled to a conformally invariant scalar field in the de Sitter invariant vacuum for an instance, if the atom is freely falling, it would spontaneously excite as if it were immersed in a thermal bath at 
$T_{GH}$;  while if it is static, the temperature of the thermal bath becomes $T_S=\sqrt{T_{GH}^2+T_{U}^2}$, where $T_{U}$ is the Unruh temperature associated with the inherent acceleration of the static atom~\cite{Zhu08}.

The aforementioned studies~\cite{Gibbons77,Deser97,Zhu08,Zhou10,Tian13,Oshita14} show that as far as one detector is concerned, de Sitter spacetime is indistinguishable from a thermal bath in Minkowski space in terms of the radiative properties of the detector such as the transition rates and the Lamb shift. 
Questions then arise as to what  will happen if two spatially separated detectors are considered and whether de Sitter spacetime is still equivalent to a thermal bath in Minkowski spacetime in terms of physical traits  of  the two-detector system.  The inter-detector interaction induced by  vacuum fluctuations of fields the detectors are coupled  to is such a trait, which can then be exploited to reveal the nature of de Sitter spacetime   in addition to the transition rate of a single detector. 
We will examine this inter-detector interaction to see if
the equivalence between the thermal bath as seen by a single detector in de Sitter spacetime and the thermal bath  in  Minkowski spacetime at temperature $T_S$ still holds,  and if not, how they differ.

\section{The setup and the approach.}
We consider that two static 
detectors labeled by $A$ and $B$ are located at the same radial but different azimuthal coordinates in four-dimensional de Sitter spacetime which can be represented as the hyperboloid
\bea
z^2_0-z^2_1-z^2_2-z^2_3-z^2_4=-\alpha^2
\eea
embedded in five dimensional Minkowski space with the metric
\bea
ds^2=dz^2_0-dz^2_1-dz^2_2-dz^2_3-dz^2_4\;.
\eea
Here and after, $\alpha=\sqrt{\frac{3}{\Lambda}}$. 
Applying the parametrization that
\bea
z_0&=&\sqrt{\alpha^2-r^2}\sinh{({t}/{\alpha})}\;,\nonumber\\
z_1&=&\sqrt{\alpha^2-r^2}\cosh{({t}/{\alpha})}\;,\nonumber\\
z_2&=&r\cos{\theta}\;,\\
z_3&=&r\sin{\theta}\cos{\varphi}\;,\nonumber\\
z_4&=&r\sin{\theta}\sin{\varphi}\;,\nonumber
\label{parametrization}
\eea
we obtain the following static de Sitter metric,
\bea
ds^2=\left(1-\frac{r^2}{\alpha^2}\right)dt^2-\left(1-\frac{r^2}{\alpha^2}\right)^{-1}dr^2-r^2d\theta^2-r^2\sin^2\theta d\varphi^2\;.\label{de-Sitter-metric}
\eea
Obviously, for this metric, the sphere with $r=\alpha$ is singular, and it is the so called cosmological horizon.
The detectors are then assumed to be conformally coupled to a  massless scalar field $\phi(x)$ with $x=x(t,r,\theta,\varphi)$ in the de Sitter invariant vacuum~\cite{Allen85,Polarski89}, which satisfies
\bea
\left(\nabla_{\nu}\nabla^{\nu}+\frac{1}{6} R \right)\phi(x)=0
\eea
with $R=12\alpha^{-2}$ being the scalar curvature of de Sitter spacetime.  We can expand $\phi(x)$ in terms of a complete set of field modes $f_{\mathbf{k}}$ which are solutions of the above equation 
as
\bea
\phi(x)=\int d^3{\mathbf{k}}\left[a_{\mathbf{k}}(t)f_{\mathbf{k}}(\mathbf{x})+a^{\dag}_{\mathbf{k}}(t)f^{*}_{\mathbf{k}}(\mathbf{x})\right]\label{phi}
\eea
with $a_{\mathbf{k}}$ and $a^{\dag}_{\mathbf{k}}$ being the annihilation and creation operators with momentum $\mathbf{k}$.

We now model the two detectors as two two-level atoms,  and denote the ground state and the excited state of atom $\xi(=A,B)$ by $|g_{\xi}\rangle$ and $|e_{\xi}\rangle$ and  the energy level gap by $\omega_{\xi}$. Since the radial coordinates of the two atoms are identical, the two atoms share the same proper time $\tau$. Then the interaction Hamiltonian between the atoms and the field can be described by~\footnote{Throughout the paper, we exploit the units that $\hbar=c=k_B=1$.}
\bea
H_I(\tau)=\mu R^{A}_{2}(\tau)\phi(x_{A}(\tau))+\mu R^{B}_{2}(\tau)\phi(x_{B}(\tau))\label{HI}
\eea
with $\mu$ being a very small coupling constant, $R^{\xi}_{2}=\frac{i}{2}(R_{-}^{\xi}-R_{+}^{\xi})$, $R_{-}^{\xi}=|g_{\xi}\rangle\langle e_{\xi}|$ and $R_{+}^{\xi}=|e_{\xi}\rangle\langle g_{\xi}|$.

Because each atom is perturbed by the fluctuating field $\phi(x)$ in the vacuum, it emits a radiative field as a backreaction which then acts on the other atom and thus an interatomic interaction potential is resulted. We exploit the fourth-order DDC formalism~\cite{Dalibard82,Dalibard84} to calculate this interaction potential, which allows a distinct separation of the contributions of the vacuum fluctuations of the field and the radiation reaction of the atoms. 
This formalism was proposed by Dalibard, Dupont-Roc and Cohen-Tannoudji [the DDC formalism]~\cite{Dalibard82,Dalibard84} in an attempt to understand the dynamics of an atomic system coupled to the radiation field and gain insight into  the radiative processes in terms of  fluctuations of two interacting systems, i.e.,  a large reservoir and a small quantum system.
The DDC formalism  was then widely utilized to investigate various second-order vacuum-fluctuation-induced effects including resonant interactions~\cite{Meschede90,Jhe91,Audretsch94,Audretsch95,Holzmann95,Passante98,Tomazelli03,Rizzuto07,Zhu08,Zhou12,Rizzuto16,Zhou16}, just to name a few. Very recently, to deal with the interaction between two ground-state atoms in interaction with fluctuating scalar fields in vacuum, which is a fourth-order perturbation effect, this formalism has been generalized from the second order in its original form to the fourth order~\cite{Zhou21}.

According to  Ref.~\cite{Zhou21}, 
 the contribution of the vacuum fluctuations to the interatomic interaction potential between the two ground-state atoms [vf-contribution] is given by
\bea
\label{2vf}
(\delta E)_{vf}&=&2i\mu^4\int_{\tau_0}^{\tau}d\tau_1\int_{\tau_0}^{\tau_1}d\tau_2\int_{\tau_0}^{\tau_2}d\tau_3C^F(x_A(\tau),x_B(\tau_3))
\chi^F(x_A(\tau_1),x_B(\tau_2))\chi^A(\tau,\tau_1)\nonumber\\&&\times\chi^B(\tau_2,\tau_3)+\text{$A\rightleftharpoons B$ {\it term}}\;,
\eea
while that of the radiation reaction of the atoms [rr-contribution] by
\bea
\label{2rr}
(\delta E)_{rr}
&=&2i\mu^4\int_{\tau_0}^{\tau}d\tau_1\int_{\tau_0}^{\tau_1}d\tau_2\int_{\tau_0}^{\tau_2}d\tau_3\chi^F(x_A(\tau),x_B(\tau_3))\chi^F(x_A(\tau_1),x_B(\tau_2))C^A(\tau,\tau_1)\chi^B(\tau_2,\tau_3)\nonumber\\
&+&2i\mu^4\int_{\tau_0}^{\tau}d\tau_1\int_{\tau_0}^{\tau_1}d\tau_2\int_{\tau_0}^{\tau_2}d\tau_3\chi^F(x_A(\tau_1),x_B(\tau_3))\chi^F(x_B(\tau_2),x_A(\tau))C^A(\tau,\tau_1)\chi^B(\tau_2,\tau_3)\nonumber\\
&+&2i\mu^4\int_{\tau_0}^{\tau}d\tau_1\int_{\tau_0}^{\tau_1}d\tau_2\int_{\tau_0}^{\tau_2}d\tau_3\chi^F(x_A(\tau_3),x_B(\tau_2))\chi^F(x_B(\tau_1),x_A(\tau))C^A(\tau,\tau_3)\chi^B(\tau_1,\tau_2)\nonumber\\
&+&2i\mu^4\int_{\tau_0}^{\tau}d\tau_1\int_{\tau_0}^{\tau}d\tau_2\int_{\tau_0}^{\tau_2}d\tau_3\chi^F(x_A(\tau_2),x_B(\tau_3))\chi^F(x_A(\tau),x_B(\tau_1))\chi^A(\tau,\tau_2)C^B(\tau_1,\tau_3)\nonumber\\
&+&2i\mu^4\int_{\tau_0}^{\tau}d\tau_1\int_{\tau_0}^{\tau_1}d\tau_2\int_{\tau_0}^{\tau}d\tau_3C^F(x_B(\tau_2),x_A(\tau_3))\chi^F(x_B(\tau_1),x_A(\tau))\chi^A(\tau_3,\tau)\chi^B(\tau_1,\tau_2)\nonumber\\
&+&2i\mu^4\int_{\tau_0}^{\tau}d\tau_1\int_{\tau_0}^{\tau_1}d\tau_2\int_{\tau_0}^{\tau_1}d\tau_3\chi^F(x_A(\tau),x_B(\tau_3))\chi^F(x_A(\tau_1),x_B(\tau_2))C^A(\tau,\tau_1)\chi^B(\tau_3,\tau_2)\nonumber\\
&+&\text{$A\rightleftharpoons B$ {\it terms}}\;,
\eea
In the above two equations, $C^{\xi}$ and $\chi^{\xi}$ are the symmetric and antisymmetric statistical functions of the atoms defined by
\bea
C^{\xi}(\tau,\tau')&\equiv&\frac{1}{2}\langle g_{\xi}|\{R^{\xi,f}_{2}(\tau),R^{\xi,f}_{2}(\tau')\}|g_{\xi}\rangle\;,\label{C-atom}\\
\chi^{\xi}(\tau,\tau')&\equiv&\frac{1}{2}\langle g_{\xi}|[R^{\xi,f}_{2}(\tau),R^{\xi,f}_{2}(\tau')]|g_{\xi}\rangle\label{Chi-atom}
\eea
with
\bea\label{R2f}
R^{\xi,f}_{2}(\tau)=\frac{i}{2}\left[R^{\xi}_{-}(\tau_0)e^{-i\omega_{\xi}(\tau-\tau_0)}-R^{\xi}_{+}(\tau_0)e^{i\omega_{\xi}(\tau-\tau_0)}\right]
\eea
the free part of the atomic operator $R^{\xi}_{2}(\tau)$, and $C^F$ and $\chi^F$ are the symmetric correlation function and the linear susceptibility of the scalar field defined by
\bea
C^F(x_A(\tau),x_B(\tau'))&\equiv&\frac{1}{2}\langle 0|\{\phi^{f}(x_A(\tau)),\phi^{f}(x_B(\tau'))\}|0\rangle\;,\quad\label{fieldC}\\
\chi^F(x_A(\tau),x_B(\tau'))&\equiv&\frac{1}{2}\langle 0|[\phi^{f}(x_A(\tau)),\phi^{f}(x_B(\tau'))]|0\rangle\label{fieldChi}
\eea
with $\phi^f(x)$ being
the free part of the scalar field operator not including the radiative field of the atoms and $|0\rangle$ the de Sitter-invariant vacuum state~\cite{Allen85}. 
Adding up Eqs.~(\ref{2vf}) and (\ref{2rr}), we then obtain the total interaction potential, $(\delta E)_{tot}=(\delta E)_{vf}+(\delta E)_{rr}$.

\section{Interatomic interaction potential in de Sitter spacetime.}
To calculate the vf- and rr-contribution with Eqs.~(\ref{2vf}) and (\ref{2rr}), we must first compute the symmetric correlation function $C^F$ and the linear susceptibility  $\chi^F$ of the field along the trajectories of the atoms, which, with an appropriate choice of coordinates, can be described by $x_A(\tau)=(t_A(\tau),r_0,\frac{\pi}{2},\varphi_A)$ and $x_B(\tau)=(t_B(\tau),r_0,\frac{\pi}{2},\varphi_B)$ 
with $\varphi_{\xi}\in[0,\pi]$ and $\varphi_B>\varphi_A$. Combining these trajectories, the metric Eq.~(\ref{parametrization}) and the following two-point correlation function of the scalar field at two arbitrary points $x$ and $x'$~\cite{Polarski89}
\bea
\langle 0|\phi^{f}(x)\phi^{f}(x')|0\rangle=-\frac{1}{4\pi^2}\Big[(z_0-z'_0-i\epsilon)^2-\sum_{i=1}^4(z_i-z'_i)^2\Big]^{-1}
\eea
with Eqs.~(\ref{fieldC}) and (\ref{fieldChi}), we obtain
\bea
C^F(x_A(\tau),x_B(\tau'))=\frac{1}{8\pi^2}\int_{0}^{\infty}d\omega\frac{\sin\left[\omega F(r_0,\varphi_0)\right]}{H(r_0,\varphi_0)}
\coth\big(\pi\omega\alpha\sqrt{g_{00}(r_0)}\big)\big(e^{-i\omega\Delta\tau}+e^{i\omega\Delta\tau}\big),\quad\label{cf}\\
\chi^F(x_A(\tau),x_B(\tau'))=\frac{1}{8\pi^2}\int_{0}^{\infty}d\omega
\frac{\sin\left[\omega F(r_0,\varphi_0)\right]}{H(r_0,\varphi_0)}\left(e^{-i\omega\Delta\tau}-e^{i\omega\Delta\tau}\right)\;,\quad\quad\quad\quad\;\;\label{chif}
\eea
where $\Delta\tau=\tau-\tau'$, $\varphi_0=\varphi_B-\varphi_A$, $g_{00}(r_0)=1-r_0^2/\alpha^2$, and
\bea
F(r_0,\varphi_0)&\equiv&2\alpha\sqrt{g_{00}(r_0)}\sinh^{-1}\biggl(\frac{r_0\sin(\varphi_0/2)}{\alpha\sqrt{g_{00}(r_0)}}\biggr)\;,\\
H(r_0,\varphi_0)&\equiv&2r_0\sin(\varphi_0/2)\sqrt{1+\frac{r_0^2\sin^2(\varphi_0/2)}{\alpha^2g_{00}(r_0)}}\;.
\eea Here,
in expressing  $C^F(x_A(\tau),x_B(\tau'))$ and $\chi^F(x_A(\tau),x_B(\tau'))$ as an integration over $\omega$, we have used the method of the Fourier transform.  For brevity, we next abbreviate $F(r_0,\varphi_0)$ and $H(r_0,\varphi_0)$ which both have the dimension of length as $F$ and $H$, and $g_{00}(r_0)$ which is dimensionless as $g_{00}$.

Putting Eq.~(\ref{R2f}) in Eqs.~(\ref{C-atom}) and (\ref{Chi-atom}), we can further simplify the two statistical functions of the atoms. 
Then, inserting them and Eqs.~(\ref{cf}) and (\ref{chif}) into Eqs.~(\ref{2vf}) and (\ref{2rr}),  and performing the triple integrations with respect to $\tau_1$, $\tau_2$ and $\tau_3$ for an infinitely long time interval, i.e., $(\tau-\tau_0)\rightarrow\infty$, we can express the respective vf- and rr-contribution to the interaction potential as
\bea
(\delta E)_{vf}=-\frac{\mu^4\omega_A\omega_B}{64\pi^4H^2}\int_0^{\infty}d\omega_1\int_0^{\infty}d\omega_2
\frac{\omega_2\sin(\omega_1F)\sin(\omega_2F)
\coth{\big(\pi\omega_1\alpha\sqrt{g_{00}}\big)}}{(\omega_1^2-\omega_A^2)(\omega_1^2-\omega_B^2)(\omega_2^2-\omega_1^2)}\quad
\label{vf}
\eea
and
\bea
(\delta E)_{rr}&=&-\frac{\mu^4}{32\pi^4H^2}\int_0^{\infty}d\omega_1\int_0^{\infty}d\omega_2
\frac{\omega_2\sin(\omega_1F)\sin(\omega_2F)}{\omega_2^2-\omega_1^2}
\biggl[\frac{\omega_1+\omega_A+\omega_B}{(\omega_1+\omega_A)(\omega_1+\omega_B)(\omega_A+\omega_B)}\nonumber\\&&
-\frac{\omega_A\omega_B\big(1-\frac{1}{2}\coth{\big(\pi\omega_1\alpha\sqrt{g_{00}}\big)}\big)}{(\omega_1^2-\omega_A^2)(\omega_1^2-\omega_B^2)}\biggl]\;.
\label{rr}
\eea
Finally, adding up the above two equations, we arrive at the following total interaction potential after some simplifications,
\bea
(\delta E)_{tot}&=&-\frac{\mu^4\omega_A\omega_B}{128\pi^3H^2}\int_{0}^{\infty}du\biggl[\frac{e^{-2uF}}{(u^2+\omega_A^2)(u^2+\omega_B^2)}
+\frac{2\sin(2uF)}{(u^2-\omega_A^2)(u^2-\omega_B^2)(e^{2\pi u\alpha\sqrt{g_{00}}}-1)}\biggl]\;.\nn\\
\label{de}
\eea
We next study how the interaction potential behaves as the physical separation varies in some special regimes. Let us note that
 the physical interatomic separation, i.e., the length of the geodesic connecting  two spatial  points $(r_0,\frac{\pi}{2},\varphi_A)$ and $(r_0,\frac{\pi}{2},\varphi_B)$, is given by
\bea
\rho\equiv\rho(r_0,\varphi_0)=2\sqrt{\alpha^2-r_0^2 \cos^2\left(\frac{\varphi_0}{2}\right)}\tan^{-1}\biggl(\frac{r_0\sin\left(\frac{\varphi_0}{2}\right)}{\alpha\sqrt{g_{00}}}\biggr)\label{rho}
\eea
with $\varphi_0=\varphi_B-\varphi_A$ [see the Appendix].

For an $\alpha$ approaching infinity, both $F$ and $H$ are equal to $\rho=2r_0\sin(\varphi_0/2)$, and the second term on the right of Eq.~(\ref{de}) vanishes. As a result, the total interaction potential Eq.~(\ref{de}) reduces to
\beq
(\delta E)^M_{tot}=-\frac{\mu^4\omega_A\omega_B}{128\pi^3\rho^2}\int_{0}^{\infty}du\frac{e^{-2u\rho}}{(u^2+\omega_A^2)(u^2+\omega_B^2)}
\label{Minkowski-vacuum-inter-potential}
\eeq
which is exactly the interaction potential of two static atoms at a  separation $\rho$ in the flat Minkowski spacetime~\cite{Cheng-thermal}. This is physically expected, as the de Sitter metric Eq.~(\ref{de-Sitter-metric}) reduces to the Minkowski metric when $\alpha\rightarrow\infty$.

However, for a finite $\alpha$,  there does not exist a simple relation between $F$ ($H$) and the physical inter-detector separation $\rho$. In order to find how the interaction potential varies with $\rho$, let us now
analyze the behavior of the interaction potential in two special regions where an explicit relation between $F$ ($H$) and $\rho$ exists, i.e., the region  far away from or very close to the cosmological horizon, where  $r_0\ll\alpha$  or $r_0\lesssim\alpha$ respectively. For simplicity, we next assume that the two atoms are identical with the same transition frequency $\omega_0$.

\subsection{Interaction potential of two atoms far away from the cosmological horizon.}
When the two atoms are located very far away from the cosmological horizon, i.e., when $r_0\ll\alpha$, 
both $F$ and $H$ are approximated by $\rho\approx2r_0\sin(\varphi_0/2)$ and the temperature as observed by a single detector $T_S\sim T_{GH}\equiv\frac{1}{2\pi\alpha}$. A thermal wavelength $\beta_{GH}=T^{-1}_{GH}$ can then be defined. So $r_0\ll\alpha$ means $r_0\ll\beta_{GH}$. Note that there is another characteristic length for the problem under consideration, i.e., the transition wavelength of the atoms, $\lambda=2\pi\omega_0^{-1}$. 

We then find that when  $\rho\ll\lambda\ll\beta_{GH}$, the vf-contribution to the interatomic interaction potential Eq.~(\ref{vf}) can be approximated by
\bea
\label{short1vf}
(\delta E)_{vf}\approx \frac{\mu^4}{256\pi^3\rho}-\frac{\mu^4T_{GH}^2}{384\pi\omega_0^2 \rho}\biggl(1-\frac{1}{4}\rho^2\omega_0^2\biggr)\;,
\eea
which results in a repulsive interaction force, and the rr-contribution Eq.~(\ref{rr}) by
\bea
\label{short1rr}
(\delta E)_{rr}\approx-\frac{\mu^4}{512\pi^2\omega_0\rho^2}+\frac{\mu^4}{256\pi^3\rho}-\frac{\mu^4 T_{GH}^2}{384\pi \omega_0^2\rho}\biggl(1-\frac{1}{4}\rho^2\omega_0^2\biggr)\;,
\eea
which leads to an attractive interaction force  much greater than the vf-contribution.

Adding Eqs.~(\ref{short1vf}) and (\ref{short1rr}) up, we obtain the following total interaction potential,
\bea
\label{short1}
(\delta E)_{tot}\approx
-\frac{\mu^4}{512\pi^2\omega_0\rho^2}+\frac{\mu^4}{128\pi^3\rho}
-\frac{\mu^4 T_{GH}^2}{192\pi \omega_0^2 \rho}\biggl(1-\frac{1}{4}\rho^2\omega_0^2\biggr)\;.
\eea
This interaction potential is dominated by the first term coming from the rr-contribution, which is negative and proportional to $\rho^{-2}$, and it implies an attractive force  between the two atoms  behaving as $\rho^{-3}$. As compared with the interaction potential in a thermal bath  in  Minkowski spacetime~\cite{Cheng-thermal}, 
the first two terms on the right of Eq.~(\ref{short1}) which are leading over the third term are identical to those in the latter case. However, the third term which is proportional to $T_{GH}^2$ and thus manifests the thermal effects of  de Sitter spacetime on the interatomic interaction, displays a new separation-dependence due to the existence of the extra factor $\left(1-\frac{1}{4}\rho^2\omega_0^2\right)$ which is slightly smaller than unity, meaning an interaction potential slightly smaller than that in the Minkowski thermal bath. This distinction however small suggests that the thermal nature of  de Sitter space is basically different from that of a thermal bath at temperature $T_S$  in  Minkowski spacetime as is revealed by a single detector.

\subsection{Interaction potential of two atoms near the cosmological horizon. }

When the two atoms are very close to the cosmological horizon, i.e., $r_0\lesssim\alpha$, $\rho\sim\pi\alpha\sin(\varphi_0/2)$,  and
$T_S\sim T_{U}\equiv\frac{a(r_0)}{2\pi}$\label{Tnear}  with $a(r_0)\equiv\frac{r_0}{\alpha\sqrt{\alpha^2-r_0^2}}$
which is very large. Accordingly, the characteristic wavelength $\beta_{U}=T_U^{-1}$ is extremely small. If we  further assume $\rho\gg\lambda$,  then for two atoms located very close to the cosmological horizon, we have $\beta_{U}\ll\lambda\ll\rho$. In this case, the vacuum fluctuations and the radiation reaction of the atoms yield almost equally important contributions to the interatomic interaction potential, which are both approximated by
\bea
\label{long3vf}
(\delta E)_{vf}\approx(\delta E)_{rr}\approx\frac{\mu^4}{4096\omega^2_0 T_{U}\rho^4\ln(4T_{U}\rho)}\;,
\eea
and thus
\bea
\label{long3}
(\delta E)_{tot}\approx\frac{\mu^4}{2048\omega_0^2T_{U}\rho^4\ln(4T_{U}\rho)}\;,
\eea
which leads to a repulsive force between the two atoms. Eqs.~(\ref{long3vf}) and (\ref{long3}) show a completely new behavior of $\sim[T_{U}\rho^4\ln(4T_{U}\rho)]^{-1}$ for the interaction potential in sharp contrast to that of two atoms located very far away from the cosmological horizon [refer to Eqs.~(\ref{short1vf})-(\ref{short1})]. More importantly, this behavior also deviates dramatically from its counterpart in the Minkowski thermal bath, as the vf- and rr-contribution and thus the total interaction potential in the latter case oscillate obviously with the interatomic separation and the interaction force between the two atoms can be either attractive or repulsive and even be null~\cite{Cheng-thermal}, while in the present case the interaction potential decays monotonically with the interatomic separation and it generates a repulsive force.


As this novel behavior emerges 
in the regime when the effect of proper acceleration of the static atoms with respect to the locally inertial ones dominates, one may wonder whether the thermal  nature of de Sitter space as revealed by the interaction potential may approximate to that of the Unruh thermal bath associated with the proper acceleration of the static atoms. To answer this question, let us
 make a comparison of the result in the present case and that in the case of two atoms in synchronous uniform acceleration perpendicular to the  interatomic separation in  Minkowski spacetime.  We then see that the behavior of the interatomic interaction we found in the present paper clearly differs from $\sim(T_U\rho^4)^{-1}$ for two uniformly accelerated atoms in  Minkowski spacetime~\cite{Zhouacc}.
So, de Sitter spacetime as seen by two detectors in terms of the inter-detector interaction is also distinctive from an Unruh thermal bath.

Therefore, the thermal nature of de Sitter space is intrinsically different both from the thermal bath in Minkowski space and the Unruh thermal bath felt by non-inertial observers, and in principle one can tell from the behavior of the interatomic  interaction  as the physical inter-detector separation varies whether the detectors are in de Sitter spacetime or in a thermal bath in Minkowski spacetime or even the detectors are uniformly accelerating themselves with respect to an inertial frame. In this sense, de Sitter spacetime is unique on its own right in regard to its thermal nature, and it is neither equivalent to a thermal bath in Minkowski spacetime nor to an Unruh thermal bath as seen by non-inertial observers. Finally, it is worth pointing out that  this remarkable thermal character of de Sitter spacetime may also be revealed by examining the entanglement dynamics of a pair of detectors~\cite{Steeg09,Salton15}. However, it cannot be disclosed via the resonance interaction between two detectors [atoms] in one of the maximally entangled states, i.e., the symmetric or antisymmetric entangled state, although more than one detector is also involved~\cite{Tian16}, as the interaction energy is now insusceptible to the state of a field [thermal or nonthermal]~\cite{Zhou16}.

\section{Summary.}

In this paper, we have studied the interaction potential of two static detectors [modeled as  ground-state two-level atoms] at the same radial but different azimuthal coordinates which are conformally coupled to a scalar field in the de Sitter invariant vacuum. We discover that de Sitter spacetime in regard to  its thermal nature as disclosed by the inter-detector interaction induced by coupling with the fluctuating vacuum fields is remarkably different both from  a thermal bath in Minkowski spacetime and an Unruh thermal bath as seen by non-inertial observers. In this sense, the thermal nature of de Sitter space is unique in its own right. In principle, one can tell from the behavior of the inter-detector interaction as the physical inter-detector separation varies whether the detectors are in de Sitter spacetime or in a thermal bath in Minkowski spacetime or even the detectors are uniformly accelerating themselves  with respect to an inertial frame.

\begin{acknowledgments}
This work was supported in part by the NSFC under Grants No. 11690034, No. 11875172, No. 12075084, No. 12047551, and No. 12105061; and the K.C. Wong Magna Fund in Ningbo University.
\end{acknowledgments}

\appendix

\section{Derivation of interatomic proper separation.}\label{proper-separation}

In this Appendix, we give a derivation of the length of the geodesic connecting  two atoms at $(r_0,\frac{\pi}{2},\varphi_A)$ and $(r_0,\frac{\pi}{2},\varphi_B)$ with $\varphi_0=\varphi_B-\varphi_A>0$ in de Sitter spacetime.

With the metric Eq.~(\ref{de-Sitter-metric}), it is easy to deduce the Lagrangian of a particle freely falling in de Sitter spacetime,
\bea
\mathcal{L}=\frac{1}{2}\left[\left(1-\frac{r^2}{\alpha^2}\right)\dot t^2-\left(1-\frac{r^2}{\alpha^2}\right)^{-1}\dot r^2-r^2\dot\theta^2-r^2\sin^2\theta \dot\varphi^2\right]\;,
\label{Lagrangian}
\eea
where a dot over the coordinate variables represents the derivative with respect to an arbitrary affine parameter $\lambda$ ($\neq\tau$ for photons).
Substituting this Lagrangian into the equation of motion of free particles
\bea
\frac{d}{d\lambda}\frac{\partial \mathcal{L}}{\partial \dot x^{\nu}}-\frac{\partial \mathcal{L}}{\partial x^{\nu}}=0
\eea
with $x^{\nu}=(t,r,\theta,\varphi)$, we obtain
\bea
\frac{d}{d\lambda}(r^2\dot\theta)-r^2\sin\theta\cos\theta\dot\varphi^2&=&0\;,\ \qquad \ \text{for\ } x^{\nu}=\theta\label{theta}\\
\left(1-\frac{r^2}{\alpha^2}\right)\dot t&=&E_0\;,\qquad \text{for\ } x^{\nu}=t\label{E0}\\
r^2\sin^2\theta \dot\varphi&=&L_{\varphi}\;,\qquad \text{for\ } x^{\nu}=\varphi\label{Lphi}
\eea
where $E_0$ and $L_{\varphi}$ denote the constant energy and angular momentum of the particle, respectively.


For photons travelling along a null geodesic on the plane $\theta_0=\frac{\pi}{2}$, we have $ds^2=0$ and $\dot\theta_0=0$, and thus
\beq
\left(1-\frac{r^2}{\alpha^2}\right)\left(\frac{dt}{d\lambda}\right)^2
-\left(1-\frac{r^2}{\alpha^2}\right)^{-1}\left(\frac{dr}{d\lambda}\right)^2
-r^2\left(\frac{d\varphi}{d\lambda}\right)^2=0.
\label{photon-raidal-equation}
\eeq
Combining Eqs.~(\ref{E0}) and (\ref{Lphi}) with the above equation, we arrive at
\bea
dr=d\varphi\sqrt{\frac{E^2_0r^4}{L^2_{\varphi}}-r^2\biggl(1-\frac{r^2}{\alpha^2}\biggr)}\;,
\eea
which  gives rise to the trajectory after integration
\bea\label{rphi}
r(\varphi)=r_0\cos{\left(\frac{\varphi_0}{2}\right)}\frac{1}{\cos{\left|\varphi-\frac{\varphi_0}{2}\right|}}
\eea
for the geodesic connecting  two points $(r_0,\frac{\pi}{2},\varphi_A)$ and $(r_0,\frac{\pi}{2},\varphi_B)$ with $\varphi_B-\varphi_A=\varphi_0$.

Finally, applying Eq. (\ref{rphi}) in the following expression of the proper separation of the two atoms,
\bea
\rho(r_0,\varphi_0)&=&\int_{\varphi_A}^{\varphi_{B}}d\varphi\sqrt{\left(1-\frac{r^2(\varphi)}{\alpha^2}\right)^{-1}\left(\frac{dr(\varphi)}{d\varphi}\right)^2+r^2(\varphi)}\;,
\eea
we obtain
\bea
\rho(r_0,\varphi_0)=2\sqrt{\alpha^2-r_0^2 \cos^2\left(\frac{\varphi_0}{2}\right)}\tan^{-1}\biggl(\frac{r_0\sin\left(\frac{\varphi_0}{2}\right)}{\alpha\sqrt{g_{00}}}\biggr)\;,
\eea
which is accurately Eq.~(\ref{rho}).


\begin{thebibliography}{}
\bibitem{Strominger01} A. Strominger, J. High Energy Phys. {\bf 10}, 34 (2001); {\bf 11}, 49 (2001).
\bibitem{Gibbons77} G. W. Gibbons and S. Hawking, 
                    Phys. Rev. D {\bf 15}, 2738 (1977).
\bibitem{Deser97} S. Deser and O. Levin, 
                  Class. Quantum Grav. {\bf 14}, L163 (1997).
\bibitem{Yu11} H. Yu, 
               Phys. Rev. Lett. {\bf 106}, 061101 (2011).
\bibitem{Narnhofere96} H. Narnhofere, I. Peter, and W. Thirring, 
                       Int. J. Mod. Phys. B {\bf 10}, 1507 (1996).
\bibitem{Hutasoit09} J. A. Hutasoit, S. P. Kumar, and J. Rafferty, 
                     J. High Energy Phys. {\bf 04}, 063 (2009).
\bibitem{Henry10} S.-H. Henry Tye, D. Wohns, and Y. Zhang, 
                  Int. J. Mod. Phys. A {\bf 25}, 1019 (2010).
\bibitem{Zhu08} Z. Zhu and H. Yu, 
                {J. High Energy Phys. {\bf 02}, 33 (2008)}.
\bibitem{Zhou10} W. Zhou and H. Yu, 
                 {Phys. Rev. D {\bf 82}, 124067 (2010)}.
\bibitem{Tian13} Z. Tian and J. Jing, 
           J. High Energy Phys. {\bf 04}, 109 (2013).
\bibitem{Oshita14} N. Oshita, K. Yamamoto, and S. Zhang, 
                   Phys. Rev. D {\bf 89}, 124028 (2014).
\bibitem{Allen85} B. Allen, 
                 {Phys. Rev. D {\bf 32}, 3136 (1985)}.
\bibitem{Polarski89} D. Polarski, 
                     Class. Quantum Grav. {\bf 6}, 717 (1989).
\bibitem{Dalibard82} J. Dalibard, J. Dupont-Roc, and C. Cohen-Tannoudji, 
                    {J. Phys. France {\bf 43}, 1617 (1982)}.
\bibitem{Dalibard84} J. Dalibard, J. Dupont-Roc, and C. Cohen-Tannoudji, 
                    {J. Phys. France {\bf 45}, 637 (1984)}.
\bibitem{Meschede90} D. Meschede, W. Jhe, and E. A. Hinds, 
                     {Phys. Rev. A} {\bf 41}, 1587 (1990).
\bibitem{Jhe91} W. Jhe, 
               {{Phys. Rev. A} {\bf 43}, 5795 (1991)}.
\bibitem{Audretsch94} J. Audretsch and R. M$\ddot{\text{u}}$ller, 
                     {Phys. Rev. A {\bf 50}, 1755 (1994)}.
\bibitem{Audretsch95} J. Audretsch and R. M$\ddot{\text{u}}$ller, 
                     {Phys. Rev. A {\bf 52}, 629 (1995)}.
\bibitem{Holzmann95} J. Audretsch, R. M$\ddot{\text{u}}$ller, and M. Holzmann, 
                    {Class. Quantum Grav. {\bf 12}, 2927 (1995)}.
\bibitem{Passante98} R. Passante, 
                    {Phys. Rev. A {\bf 57}, 1590 (1998)}.
\bibitem{Tomazelli03} J. L. Tomazelli and L. C. Costa, 
                      Int. J. Mod. Phys. A {\bf 18}, 1079 (2003).
\bibitem{Rizzuto07} L. Rizzuto, 
                   {Phys. Rev. A {\bf 76}, 062114 (2007)}.
\bibitem{Zhou12} W. Zhou and H. Yu, 
                 Phys. Rev. A {\bf 86}, 033841 (2012).
\bibitem{Rizzuto16} L. Rizzuto, M. Lattuca, J. Marino, A. Noto, S. Spagnolo, W. Zhou, and R. Passante, 
                   {Phys. Rev. A {\bf 94}, 012121 (2016)}.
\bibitem{Zhou16} W. Zhou, R. Passante, and L. Rizzuto, 
                 Phys. Rev. D {\bf 94}, 105025 (2016).
\bibitem{Zhou21} W. Zhou, S. Cheng, and H. Yu, 
                {Phys. Rev. A {\bf 103}, 012227 (2021)}.
\bibitem{Cheng-thermal} S. Cheng, W. Zhou, and H. Yu, 
                        {Commun. Theor. Phys. {\bf 74}, 125103 (2022)}.
\bibitem{Zhouacc} W. Zhou, S. Cheng, and H. Yu, 
                  {Phys. Lett. B {\bf 834}, 137440 (2022)}.
\bibitem{Steeg09} G. V. Steeg and N. C. Menicucci, 
                  Phys. Rev. D {\bf 79}, 044027 (2009).
\bibitem{Salton15} G. Salton, R. B. Mann, and N. C. Menicucci, 
                   New J. Phys. {\bf 17}, 035001 (2015).
\bibitem{Tian16} Z. Tian, J. Wang, J. Jing, and A. Dragan, 
                 Sci. Rep. {\bf 6}, 35222 (2016).

\end{thebibliography}
\end{document}